# Modeling of two-dimensional DNA display

Ana-Maria FLORESCU [(1)], Marc JOYEUX [(1)] and Bénédicte LAFAY [(2)]

(1) Laboratoire de Spectrométrie Physique (CNRS UMR5588), Université Joseph Fourier Grenoble 1, BP 87, 38402 St Martin d'Hères, France

(2) Laboratoire Ampère (CNRS UMR5005), Université de Lyon, Ecole Centrale de Lyon, 36 avenue Guy de Collongue, 69134 Ecully, France

Corresponding author :

Marc JOYEUX

Laboratoire de Spectrométrie Physique (CNRS UMR5588)

Université Joseph Fourier Grenoble 1

BP 87

38402 St Martin d'Hères

France

Email : Marc.Joyeux@ujf-grenoble.fr

Phone : (+33) 476 51 47 51

Fax : (+33) 476 63 54 95

**Abstract :** 2D display is a fast and economical way of visualizing polymorphism and comparing genomes, which is based on the separation of DNA fragments in two steps, according first to their size and then to their sequence composition. In this paper, we present an exhaustive study of the numerical issues associated with a model aimed at predicting the final absolute locations of DNA fragments in 2D display experiments. We show that simple expressions for the mobility of DNA fragments in both dimensions allow one to reproduce experimental final absolute locations to better than experimental uncertainties. On the other hand, our simulations also point out that the results of 2D display experiments are not sufficient to determine the best set of parameters for the modeling of fragments separation in the second dimension and that additional detailed measurements of the mobility of a few sequences are necessary to achieve this goal. We hope that this work will help in establishing simulations as a powerful tool to optimize experimental conditions without having to perform a large number of preliminary experiments and to estimate whether 2D DNA display is suited to identify a mutation or a genetic difference that is expected to exist between the genomes of closely related organisms.

## 1 - Introduction

Two-dimensional (2D) DNA display was first described by Fisher and Lerman [1-3]. This is an electrophoresis technique, which consists in separating DNA fragments in two steps, according first to their size and then to their sequence composition. The first step uses traditional slab electrophoresis, for example in agarose or polyacrylamide gels. Collisions between DNA and the gel reduce the mobility of DNA fragments, so that the gel acts as a sieve and the electrophoretic mobility becomes size-dependent, with smaller molecules generally going faster than large ones [4]. In the second dimension, fragments of identical length are separated on the basis of their sequence composition, thanks to a gradient of either temperature (TGGE : temperature gradient gel electrophoresis) or the concentration of a chemical denaturant, e.g., a mixture of urea and formamide (DGGE : denaturing gradient gel electrophoresis), both methods being closely related (see [5,6] and below). The effective volume of denaturated regions being larger than that of double-stranded ones, the mobility of a given fragment decreases as the number of open base pairs increases. Since AT-rich regions melt at lower temperatures than GC-rich ones, GC-rich fragments usually move farther than AT-rich ones.

Although 2D DNA display has recently been applied to the comparison of the genomes of closely related bacteria [7-9], this method is still essentially empirical and simulations have only been used to a very limited extent to plan experiments and interpret results [10-12]. In particular, it has been shown only recently [12] that the final relative positions of DNA fragments in 2D display experiments can be predicted with satisfying precision using a model that combines step-by-step integration of the equation of motion of each fragment and the use of the open source program Meltsim [13] to estimate the number of open base pairs at each step of the DGGE phase. This kind of simulations will therefore certainly develop further, in order to help optimize experimental conditions (denaturing gradient range, electrophoresis duration, etc) without having to perform a large number of tedious preliminary experiments, and to predict whether 2D DNA display is a convenient tool to identify a given mutation or a genetic difference that is expected to exist between the genomes of closely related organisms.

The goal of this paper is to extend the results presented in [12] along several lines. First, these results were obtained by modifying one parameter in existing formulae for the mobility of the DNA fragments during each phase of the 2D display process [5,14,15]. The point is that the formula, which was used to estimate the mobility of the fragments during electrophoresis along the first dimension, is rather complex and involves many parameters [14,15]. It is therefore not ideally suited for fitting purposes. We will show that the corresponding procedure can be greatly simplified without deteriorating the quality of predictions. Moreover, calculations in [12] disregarded absolute positions and considered only relative ones. This may be dangerous, because the positions of *all* fragments will be badly predicted if, by lack of chance, the displacement of one of the two fragments that serve as a basis to define relative positions is badly modeled. In addition, while it is indisputable that traditional electrophoresis along the first dimension is a process that is linear with respect to both time and distance, which allows one to use relative instead of absolute positions, this is no longer the case for the second dimension : denaturation is indeed a very abrupt process, so that use of relative coordinates becomes rather questionable. We will show in this paper that it is possible to handle absolute rather than relative positions, provided that one modifies the relation, which is used to estimate the equivalent temperature at a given position of the denaturing gradient gel, i.e., for a given concentration of the chemical denaturant [6,16,17]. Last but not least, we adjusted all the parameters of the model, in order to get a global view of its capabilities and limitations. It will in particular be shown that, while the model is capable of predicting the final positions in the second dimension with an average error smaller than experimental uncertainties, the number of base pairs that are open at the end of the electrophoresis experiment can nevertheless not be estimated accurately.

The remainder of this article is organized as follows. The simulation procedure is sketched in Sect. 2. Simulation of the separation according to size (first dimension) and sequence composition (second dimension) is next discussed in Sect. 3. We finally conclude in Sect. 4.

**2 – Materials and methods**

According to the definition of mobility, the position $y$ of sequence $s$ at time $t$ in a constant electric field $E$ satisfies

$$\frac{dy}{dt} = \mu(s, y) E \ . \tag{2.1}$$

If the mobility $\mu(s, y)$ depends uniquely on the sequence $s$ and not on position $y$, as is the case for the standard electrophoretic set-up in the first dimension, integration of Eq. (2.1) is straightforward and leads to

$$y(t) - y(0) = \mu(s) E t \ . \tag{2.2}$$

In contrast, if the mobility $\mu(s, y)$ depends on both the sequence $s$ and position $y$, as is the case for TGGE and DGGE, then Eq. (2.1) must be integrated step by step according to

$$y(t + dt) = y(t) + \mu(s, y(t)) E \, dt \ . \tag{2.3}$$

In this work, we integrated such equations of motion for the 40 fragments already discussed in [12], which were obtained from the site-specific restrictions of λ-phage genomic DNA using EcoRI, Eco47I, Eco91I, HindIII and PstI, respectively. As reported in the first columns of Table 1, the size of these fragments varies between 1929 and 23130 base pairs, and their GC content between 36.0% and 58.9%. For the first separation (according to size), we plugged the experimental values of the electric field ($E$=2 V/cm) and the total electrophoresis time ($t$=8 h) in Eq. (2.2). For the second separation (according to sequence composition), we integrated Eq. (2.3) with the experimental value $E$=7 V/cm for 44 h by steps of 7 mn and checked that results do not vary when the total integration time is increased to 80 h and the time step lowered to 1 mn. We also checked that these results are similar to those obtained with an integration time of 24 h, which coincides with the experimental duration, and concluded that DNA sequences were already stopped at the end of the electrophoresis experiments.

As will be seen in more detail in Sect. 3.2, the calculation of $\mu(s, y(t))$ during DGGE requires the estimation of the number of open base pairs of sequence $s$ at a temperature $T$ which has the same denaturation effect as the local concentration of denaturant. This was achieved as in [12] by using the open source program MeltSim [13], which is based on Poland's algorithm [18] and Fixman and Freire's speed up approximation [19]. We used the set of thermodynamic parameters of Blake and Delcourt [20] and set the positional map resolution to 1, which corresponds to the highest possible calculation precision for the number of open base pairs. The influence of the remaining free parameter of the program, namely the salt concentration [$Na^+$], will be discussed in detail in Sect. 3.2.

At last, the mobility $\mu(s, y)$ is expressed for both electrophoresis steps as a function of a certain numbers of parameters, which need to be adjusted to reproduce experimental results accurately. Eq. (2.2) and the step by step integration of Eq. (2.3) were therefore embedded in a refinement loop based on the gradient method, in order to vary these parameters so as to minimize the root mean square deviation between experimental positions and those calculated from Eqs. (2.2) and (2.3).

**3 – Results and discussion**

*3.1 – Mobility in first dimension : separation according to size*

Van Winkle, Beheshti and Rill (vWBR) [14,15] recently proposed an empirical formula, which correctly reproduces the observed mobilities of DNA fragments for a large number of experimental conditions across the three sieving regimes [4]. This formula writes

$$\frac{1}{\mu(s)} = \frac{1}{\mu_L} - \left(\frac{1}{\mu_L} - \frac{1}{\mu_S}\right) \exp\left(\frac{-N(s)}{m}\right), \qquad (3.1)$$

where $\mu_L$ and $\mu_S$ are the mobilities of infinitely large and very small fragments, $N(s)$ is the length of the investigated DNA fragment, and $m$ denotes the typical size that separates "small" from "large" sequences. Van Winkle et al [15] furthermore published the following expressions for $\mu_L$, $\mu_S$ and $m$ :

$$\begin{aligned}\mu_L &= 1.99 \times 10^{-4} \exp(-1.59\, C) \\ \mu_S &= (3.56 - 0.58\, C) \times 10^{-4} \\ m &= 7490 + 2780\, C,\end{aligned} \qquad (3.2)$$

where $\mu_L$ and $\mu_S$ are expressed in cm$^2$/(V s) and $m$ in base pairs, while $C$ denotes the agarose gel concentration in percents. The six numerical constants in Eq. (3.2), as well as the forms of the equations themselves, are expected to be valid only for the precise system investigated by van Winkle et al [15]. Still, the somewhat different experimental conditions of [12] could be accounted for by feeding in Eq. (3.2) an adjusted gel concentration $C=0.75\%$ close to the exact value $C=0.80\%$. It was indeed shown that this leads to calculated relative positions in good agreement with observed ones [12] (note, however, that absolute positions display errors larger than 1 cm). Due to the rather rigid forms of Eq. (3.2), it is however not warranted that this kind of adjustment will prove to be sufficient for experimental conditions that differ more widely from those of Ref. [15], in particular for the very popular polyacrylamide gels, and the choice of the additional parameter(s) to adjust might become rather tricky.

A very efficient alternative to bypass this numerical problem consists in adjusting directly the parameters $\mu_L$, $\mu_S$ and $m$ of Eq. (3.1) against the final absolute locations along the first dimension. One obtains $\mu_L = (0.17 \mp 0.02) \times 10^{-4}$ cm$^2$/(V s), $\mu_S = (4.53 \mp 0.03) \times 10^{-4}$ cm$^2$/(V s) and $m = 41200 \mp 6400$, which differs substantially from the values derived from Eq. (3.2) with the adjusted gel concentration $C$=0.75%, namely $\mu_L = 0.60 \times 10^{-4}$ cm$^2$/(V s), $\mu_S = 3.12 \times 10^{-4}$ cm$^2$/(V s) and $m = 9575$. Absolute positions obtained from Eq. (3.1) and the adjusted values of $\mu_L$, $\mu_S$ and $m$ are compared to observed ones in Table 1. Experimental positions correspond to the average of the positions observed in three different experiments, while the associated uncertainties were estimated by taking the standard deviations for these three experiments. Note that the results of a fourth experiment, which differ markedly from the three other ones, were discarded. It can be checked that the root mean square deviation between observed and calculated absolute positions, that is 0.05 cm, is almost four times smaller than the average experimental uncertainty, which is 0.19 cm.

*3.2 – Mobility in second dimension: separation according to sequence composition*

It appears that very few studies have addressed the question of the electrophoretic mobility of partially melted DNA sequences. To our knowledge, there is indeed only one available model [5], which is inspired from previously existing results for the mobility of branched polymers in gels. Although this model has no firm theoretical background and should be tested under a larger range of experimental conditions, several studies performed so far have reported fairly good agreement with experimental data [16,17]. According to this model, the mobility of a partially melted DNA sequence decreases exponentially with the size of the melted regions, that is

$$\mu(s,T) = \mu_0(s)\exp\left(-\frac{p(T)}{L_r}\right), \qquad (3.3)$$

where $\mu_0(s)$ is the mobility of the fragment when it is completely double-stranded, $p(T)$ is the sum of the probabilities for each base pair to be open at temperature $T$, and $L_r$ is a size parameter, which is related to the mechanism that slows down partially melted fragments and is therefore expected to depend on gel properties (concentration and pore size) and the flexibility of single-stranded DNA. Values of $L_r$ reported in the literature range from 45 to 130 base pairs [16,17]. As already mentioned in Sect. 2, we used the open source program MeltSim [13], together with the set of thermodynamic parameters of Blake and Delcourt [20], to estimate $p(T)$. The input quantities of this program are the temperature $T$, but also the salt concentration $[Na^+]$ : it is indeed well-known that the melting temperature of a sequence varies logarithmically with $[Na^+]$. It should however be stressed that MeltSim was devised to predict the denaturation behavior of DNA sequences in cells and closely related media. Since porous gels differ sensitively from such solutions, it is not obvious that salt concentration has the same effect in cells and in gels. Moreover, we do not know how the presence of other salts in the composition of the buffer affects the melting temperature of the DNA sequences. In the simulations reported below, we therefore considered the $[Na^+]$ input of the MeltSim program as a free parameter not necessarily related to the exact salt concentration in the gel.

Eq. (3.3) is sufficient to calculate the mobility of DNA fragments in TGGE experiments, that is when a temperature gradient is imposed to the gel, because the temperature $T$ at each position $y$ of the gel is known up to a certain precision. The link between the mobility $\mu(s,y)$ of Eq. (2.3) and the mobility $\mu(s,T)$ of Eq. (3.3) is therefore straightforward. This is no longer the case for DGGE experiments, where the temperature of the plate is kept uniform around 60°C and a gradient of chemical denaturant (urea+formamide) is added to the gel in order to destabilize base pairings. In this later case, the known quantity is the concentration $C_d$ of the denaturant at each position $y$ in the gel, so that estimation of the mobility $\mu(s,y)$ requires the additional knowledge of the equivalent temperature $T$, which has the same effect as a denaturant concentration $C_d$ from the point of view of the melting of DNA fragments. A linear relation was proposed in [6], namely

$$T = 57 + \frac{1}{3.2} C_d , \qquad (3.4)$$

where $C_d$ is the concentration of the standard stock solution of urea and formamide at position $y$ (expressed in % v/v) and $T$ the equivalent temperature (expressed in °C) to feed in the MeltSim program to estimate $p(T)$ at this position. As will be discussed below, we however found that Eq. (3.4) does not enable one to reproduce the absolute positions reported in [12]. We therefore replaced Eq. (3.4) by the more general linear relation

$$T = T_0 + \alpha C_d , \qquad (3.5)$$

where $T_0$ and $\alpha$ are considered as free parameters. We also took into account the very slight increase in solvent viscosity due to the gradient of denaturant by slightly adjusting the mobilities of the DNA sequences at each time step, as described in [6].

To summarize, calculation of the mobility of DNA fragments in the second dimension requires the knowledge of the numerical values of four constants, [Na$^+$], $L_r$, $T_0$ and $\alpha$. To be really complete, one should actually include $\mu_0(s)$, the mobility of fragment $s$ when it is completely double-stranded (see Eq. (3.3)), in the list of the free parameters of the model. However, several trials convinced us that this parameter is so strongly correlated to the four other ones that it is numerically impossible to let them vary simultaneously. We therefore considered that the mobility $\mu_0(s)$ that appears in Eq. (3.3) is equal to the mobility in the first dimension obtained from Eq. (3.1). We are aware that this involves a slight approximation, since the gels in the two dimensions are not identical.

We thus varied [Na$^+$], $L_r$, $T_0$ and $\alpha$ to reproduce the experimental results of [12]. These DGGE experiments were performed with 9 cm long plates and a denaturant concentration $C_d$ increasing regularly from 25% to 100% between the extremities of the plates (the total concentration of stock denaturant was computed using the protocol given by Myers et al [21] : 100% stock denaturant corresponds to 7 M urea and 40% deionized formamide). As for the first dimension, the absolute positions and uncertainties reported in Table 1 were obtained from three different experiments, while the results of a fourth experiment, which differ markedly from those of the three other ones, were discarded. We first allowed the four parameters to vary simultaneously. This resulted in the salinity [Na$^+$] decreasing below 0.001 M, which is the limit of validity of the set of thermodynamic parameters we used in the MeltSim program. In order to understand why this happens, we next performed a series of three parameters fits ($L_r$, $T_0$ and $\alpha$) at several fixed values of [Na$^+$] ranging from 0.001 M to 0.3 M. Results are shown in Figs. 1 and 2. It is seen in Fig. 1 that the root mean square error between experimental and calculated absolute positions actually remains essentially constant in the whole range 0.001-0.3 M. Furthermore, examination of Fig. 2 indicates that the adjusted values of $T_0$ and $\alpha$ vary logarithmically with [Na$^+$]. This is not really surprising because, as we already mentioned, the melting temperature of a given sequence increases logarithmically with [Na$^+$]. At last, it is seen in the top plot of Fig. 2 that the adjusted value of $L_r$ varies between 100 and 140 base pairs, which agrees with previously reported values [16,17].

Figs. 1 and 2 are however not sufficient to illustrate how broad the space of solutions is, that is how widely each parameter can be varied while still preserving a very good agreement between observed and calculated absolute positions. To get a better insight, we show in Fig. 3 the results of a series of two parameters fits, which consisted in adjusting simultaneously $T_0$ and $\alpha$ for increasing values of $L_r$ at two fixed values of [Na$^+$], namely 0.01 and 0.1 M. It can be seen in the top plot of Fig. 3 that $L_r$ can actually be varied between 30 and 220 base pairs without letting the rms error increase by more than 0.05 cm. As shown in the middle and bottom plots of Fig. 3, the adjusted values of $T_0$ and $\alpha$ vary little with $L_r$ in this range and remain close to $T_0 = 37\,°C$ and $\alpha = 0.63\,°C$ at [Na$^+$]=0.01 M and $T_0 = 58\,°C$ and $\alpha = 0.54\,°C$ at [Na$^+$]=0.1 M.

It should be clear from the examination of Figs. 1-3 that the numerical criterion is by itself not sufficient to fix unambiguously the set of parameters to use in the model and that other criteria must be taken into account. To our mind, a very sensible criterion consists in requiring that the equivalent temperature deduced from Eq. (3.5) be equal to the true temperature of the plate in the absence of chemical denaturant, that is for $C_d = 0\,\%$. This amounts to impose $T_0 = 60\,°C$ in Eq. (3.5). We therefore performed another series of two parameters fits, which consisted in adjusting simultaneously [Na$^+$] and $\alpha$ for increasing values of $L_r$ at fixed $T_0 = 60\,°C$. Results are shown in Fig. 4. Not surprisingly, the top plot again indicates that $L_r$ can be varied between 30 and 220 base pairs without letting the root mean square error increase by more than 0.05 cm. What is, however, more interesting, is that the middle and bottom plots of Fig. 4 show that the value of [Na$^+$] to feed in the MeltSim program must be chosen in the range 0.10 to 0.15 M and that $\alpha$ consequently varies in the range 0.52 to 0.55 °C. Note that this is substantially larger than the value $\alpha$=1/3.2=0.31 °C proposed in [6], but Fig. 4 unambiguously indicates that the absolute positions measured in [12] cannot be reproduced with such a low value of $\alpha$ - at least as long as one considers that $\mu_0(s)$ in Eq. (3.3) is equal to the mobility in the first dimension obtained from Eq. (3.1).

A second criterion is clearly mandatory in order to choose between the various solutions shown in Fig. 4. To our mind, this criterion should rely on the knowledge of the number of base pairs of each sequence, which are open at the end of the electrophoresis experiment. It should indeed be realized that all the solutions shown in Fig. 4 lead to the same dynamics of the fragments, i.e., the mobility and the final position of each fragment do not depend on the chosen $(L_r,[\text{Na}^+],\alpha)$ triplet, but they do *not* lead to the same denaturation properties, i.e., to the same number of open base pairs. Stated in other words, $p(T)/L_r$ remains the same for all $(L_r,[\text{Na}^+],\alpha)$ triplets, but not $p(T)$. This is clearly illustrated in Fig. 5, which shows the evolution as a function of time of the number of open base pairs (top plot) and of the position $y$ (middle plot), as well as the evolution as a function of $y$ of the mobility $\mu(s,y)$ (bottom plot), for two fragments with respective low and high GC contents and two $(L_r,[\text{Na}^+],\alpha)$ triplets with very different values of $L_r$. More precisely, the two fragments are the 2323 bps Eco91I digest of the λ-phage with 57.8% GC content and the 6555 bps Eco47I digest of the λ-phage with 38.0% GC content, while the chosen sets of parameters are $L_r=30$ bps, $[\text{Na}^+]=0.145$ M and $\alpha=0.526$ °C, and $L_r=200$ bps, $[\text{Na}^+]=0.105$ M and $\alpha=0.524$ °C. Examination of the middle and bottom plots of Fig. 5 indicates that the calculated positions and mobilities of the two fragments are very similar for the two sets of parameters. In contrast, it can be seen in the top plot that the number of base pairs that are open at the end of the electrophoresis experiment differ widely for the two sets of parameters: the set with $L_r=30$ bps predicts that about 250 base pairs are open for both fragments, while the set with $L_r=200$ bps predicts that this number is close to 1500 (note that $250/30 \approx 1500/200$). In order to fix unambiguously the correct set of parameters, which must be used to interpret electrophoresis experiments such as those reported in [12], one should therefore complement these experiments with detailed measurements of the mobility of a few sequences, as in Fig. 4 of [16]. The positions of the bumps in the evolution of mobility, which reflect the abrupt opening of large portions of the fragment, indeed reveal the correct value of $L_r$, and consequently also of [Na$^+$] and $\alpha$.

Since these additional data are not available for the experiments reported in [12], we chose the set of parameters that leads to the smallest root mean square error, that is $L_r = 100$ bps, [Na$^+$] = 0.134 M and $\alpha = 0.540$ °C, to compare calculated and experimental absolute positions in the second dimension. Results are tabulated in the four last columns of Table 1. It is stressed that the root mean square deviation between calculated and observed absolute positions (0.15 cm) is almost twice smaller than the average experimental uncertainty (0.26 cm).

### 4 – Concluding remarks

In this paper, we have presented an (hopefully) exhaustive study of the numerical issues associated with a model aimed at predicting the final absolute locations of DNA fragments in 2D display experiments. In particular, we have shown that simple expressions for the mobility of DNA fragments in both dimensions allow one to reproduce experimental final absolute locations to better than experimental uncertainties. We have furthermore pointed out that the results of 2D display experiments are not sufficient to determine the best set of parameters for the modeling of fragments separation in the second dimension and that additional detailed measurements of the mobility of a few sequences are necessary to achieve this goal.

It was mentioned in the discussion at the end of [12], that the weakest part of this model is probably Eq. (3.3), which expresses the mobility of a partially melted DNA sequence as an exponentially decreasing function of the size of the melted regions, and that the $L_r$ parameter should include some dependence on the properties of the gel (for example its concentration and the size of the pores) and the studied DNA sequences (for example their length, whether melting occurs at the extremities or inside the fragment, whether there is a single melted region or several ones, etc). We made several attempts along these lines, but all of them were unsuccessful. The reason for this is that the errors displayed in the last column of Table 1 show no particular dependence on the length of the fragments, their GC content, the distribution of the GC content inside the fragment and the number of melted regions at each temperature. This, in turn, is probably due to the fact that experimental uncertainties, which result essentially from the difficulty to control precisely the reproducibility of experimental conditions, are almost twice as large as the root mean square deviation between experimental and calculated positions. To our mind, it will not be possible (nor will it be necessary !) to improve the model discussed here and in [12] as long as experimental uncertainties will not be made substantially smaller than what can be achieved in today's experiments.

*The authors declare no conflict of interest.*

**Table 1**: Absolute coordinates of the DNA fragments in the 2D display

| Enzyme | Length (bp) | GC % | 1st dimension | | | | 2nd dimension | | | |
|---|---|---|---|---|---|---|---|---|---|---|
| | | | $y_{exp}$ (cm) | $\sigma_{exp}$ (cm) | $y_{calc}$ (cm) | $\Delta y$ (cm) | $y_{exp}$ (cm) | $\sigma_{exp}$ (cm) | $y_{calc}$ (cm) | $\Delta y$ (cm) |
| EcoRI | 21226 | 56.9 | 2.45 | 0.04 | 2.35 | 0.10 | 3.56 | 0.21 | 3.57 | -0.01 |
| | 7421 | 44.5 | 5.02 | 0.10 | 5.09 | -0.07 | 2.39 | 0.27 | 2.41 | -0.02 |
| | 5804 | 49.6 | 6.03 | 0.14 | 6.08 | -0.05 | 3.10 | 0.19 | 3.11 | 0.00 |
| | 5643 | 43.2 | 6.20 | 0.18 | 6.20 | 0.00 | 2.07 | 0.31 | 2.18 | -0.11 |
| | 4878 | 39.7 | 6.89 | 0.18 | 6.87 | 0.02 | 1.84 | 0.35 | 1.76 | 0.08 |
| | 3530 | 44.0 | 8.61 | 0.21 | 8.54 | 0.07 | 2.36 | 0.31 | 2.47 | -0.10 |
| Eco47I | 8126 | 47.8 | 4.68 | 0.10 | 4.76 | -0.08 | 2.07 | 0.32 | 2.29 | -0.22 |
| | 6555 | 38.0 | 5.56 | 0.12 | 5.57 | -0.01 | 1.81 | 0.34 | 1.75 | 0.07 |
| | 6442 | 43.7 | 5.61 | 0.12 | 5.64 | -0.03 | 2.45 | 0.29 | 2.42 | 0.03 |
| | 3676 | 47.5 | 8.35 | 0.20 | 8.32 | 0.03 | 2.70 | 0.25 | 2.96 | -0.25 |
| | 2606 | 56.4 | 10.31 | 0.21 | 10.29 | 0.02 | 3.89 | 0.18 | 3.85 | 0.04 |
| | 2555 | 56.7 | 10.45 | 0.25 | 10.41 | 0.04 | 4.04 | 0.17 | 4.00 | 0.04 |
| | 2134 | 55.3 | 11.58 | 0.27 | 11.52 | 0.06 | 3.83 | 0.21 | 3.83 | 0.01 |
| | 2005 | 57.6 | 11.87 | 0.27 | 11.92 | -0.05 | 4.30 | 0.18 | 4.00 | 0.30 |
| | 1951 | 58.5 | 12.03 | 0.28 | 12.09 | -0.06 | 4.38 | 0.17 | 4.25 | 0.13 |
| Eco91I | 8453 | 46.7 | 4.52 | 0.08 | 4.62 | -0.10 | 2.17 | 0.30 | 2.43 | -0.25 |
| | 7242 | 47.1 | 5.13 | 0.12 | 5.18 | -0.05 | 1.79 | 0.34 | 1.76 | 0.04 |
| | 6369 | 46.0 | 5.68 | 0.14 | 5.69 | -0.01 | 2.44 | 0.26 | 2.48 | -0.04 |
| | 5687 | 56.4 | 6.12 | 0.16 | 6.17 | -0.05 | 3.51 | 0.16 | 3.77 | -0.26 |
| | 4822 | 40.2 | 6.96 | 0.19 | 6.93 | 0.03 | 2.08 | 0.32 | 2.12 | -0.04 |
| | 4324 | 58.1 | 7.46 | 0.18 | 7.46 | 0.0 | 3.88 | 0.13 | 3.94 | -0.06 |
| | 3675 | 46.0 | 8.42 | 0.20 | 8.32 | 0.10 | 2.84 | 0.25 | 2.72 | 0.12 |
| | 2323 | 57.8 | 11.06 | 0.24 | 10.99 | 0.07 | 4.06 | 0.18 | 4.01 | 0.05 |
| | 1929 | 58.9 | 12.09 | 0.28 | 12.16 | -0.07 | 4.33 | 0.19 | 4.29 | 0.04 |
| HindIII | 23130 | 55.9 | 2.32 | 0.04 | 2.22 | 0.10 | 2.81 | 0.30 | 2.25 | 0.56 |
| | 9416 | 45.0 | 4.18 | 0.09 | 4.27 | -0.09 | 2.20 | 0.31 | 2.37 | -0.17 |
| | 6682 | 48.0 | 5.49 | 0.12 | 5.49 | 0.00 | 2.70 | 0.24 | 2.83 | -0.13 |
| | 4361 | 45.2 | 7.37 | 0.17 | 7.42 | -0.05 | 2.30 | 0.30 | 2.46 | -0.15 |
| | 2322 | 37.1 | 10.97 | 0.26 | 11.00 | -0.03 | 2.36 | 0.37 | 2.29 | 0.07 |
| | 2027 | 36.0 | 11.82 | 0.26 | 11.85 | -0.03 | 2.00 | 0.41 | 1.76 | 0.24 |
| PstI | 11497 | 46.8 | 3.63 | 0.06 | 3.68 | -0.05 | 2.20 | 0.32 | 2.29 | -0.09 |
| | 5077 | 44.9 | 6.69 | 0.16 | 6.68 | 0.01 | 2.32 | 0.29 | 2.50 | -0.18 |
| | 4749 | 43.8 | 7.02 | 0.19 | 7.00 | 0.02 | 2.51 | 0.26 | 2.48 | 0.04 |
| | 4507 | 36.0 | 7.31 | 0.18 | 7.26 | 0.05 | 1.86 | 0.34 | 1.77 | 0.09 |
| | 2838 | 56.6 | 9.85 | 0.24 | 9.78 | 0.07 | 4.02 | 0.16 | 4.07 | -0.05 |
| | 2560 | 53.2 | 10.39 | 0.22 | 10.40 | -0.01 | 3.74 | 0.18 | 3.76 | -0.02 |
| | 2459 | 57.7 | 10.61 | 0.24 | 10.64 | -0.03 | 4.15 | 0.17 | 4.11 | 0.04 |
| | 2443 | 54.8 | 10.69 | 0.24 | 10.68 | 0.01 | 3.84 | 0.19 | 3.82 | 0.02 |
| | 2140 | 53.1 | 11.52 | 0.27 | 11.51 | 0.01 | 3.59 | 0.24 | 3.62 | -0.03 |
| | 1986 | 58.1 | 11.92 | 0.27 | 11.98 | -0.06 | 4.14 | 0.19 | 3.98 | 0.16 |
| rms | | | | 0.19 | | 0.05 | | 0.26 | | 0.15 |

The table indicates the size (in base pairs) of each fragment, its GC content (in %), and, for each dimension, the experimental absolute position ($y_{exp}$, in cm) averaged over three experiments, the experimental uncertainty ($\sigma_{exp}$, in cm), the calculated absolute position ($y_{calc}$, in cm) and the error ($\Delta y = y_{exp} - y_{calc}$, in cm). Absolute positions in the first dimension were obtained with the expression of mobility in Eq. (3.1) and parameters $\mu_L = 0.17 \times 10^{-4}$ cm$^2$/(V s), $\mu_S = 4.53 \times 10^{-4}$ cm$^2$/(V s) and $m = 41200$. Absolute positions in the second dimension were obtained with the expression of mobility in Eq. (3.3), the expression of equivalent temperature in Eq. (3.5), and parameters $L_r = 100$ bps, $[Na^+] = 0.134$ M, $T_0 = 60$ °C and $\alpha = 0.540$ °C.

**FIGURE CAPTIONS**

**Figure 1** : Root mean square deviations (expressed in cm) between experimental and calculated absolute positions along the second dimension (DGGE experiments) for the 40 DNA sequences listed in Table 1. The three parameters $L_r$, $T_0$ and $\alpha$ were adjusted simultaneously for each fixed value of the salinity [Na$^+$].

**Figure 2** : Adjusted values of $L_r$ (top plot, units of base pairs), $T_0$ (middle plot, units of °C), and $\alpha$ (bottom plot, units of °C), as a function of the fixed salinity [Na$^+$].

**Figure 3** : Results of a series of two parameters fits, which consisted in adjusting simultaneously $T_0$ and $\alpha$ for increasing values of $L_r$ at two fixed values of [Na$^+$] (0.01 and 0.1 M). The top plot shows the root mean square error (expressed in cm) between experimental and calculated absolute positions along the second dimension (DGGE experiments) for the 40 DNA sequences listed in Table 1. The middle plot shows the evolution of $T_0$ (expressed in °C) and the bottom plot the evolution of $\alpha$ (expressed in °C).

**Figure 4** : Results of a series of two parameters fits, which consisted in adjusting simultaneously [Na$^+$] and $\alpha$ for increasing values of $L_r$ at fixed $T_0 = 60\,°C$. The top plot shows the root mean square error (expressed in cm) between experimental and calculated absolute positions along the second dimension (DGGE experiments) for the 40 DNA sequences listed in Table 1. The middle plot shows the evolution of [Na$^+$] (expressed in M) and the bottom plot the evolution of $\alpha$ (expressed in °C).

**Figure 5** : Evolution as a function of time of the number of open base pairs (top plot) and of the position $y$ (middle plot), and evolution as a function of $y$ of the mobility $\mu(s, y)$ (bottom plot), for two different fragments and two different sets of parameters. The two fragments are the 2323 bps Eco91I digest with 57.8% GC content and the 6555 bps Eco47I digest with 38.0% GC content. The two sets of parameters are $L_r = 30$ bps, $[\text{Na}^+] = 0.145$ M and $\alpha = 0.526$ °C, and $L_r = 200$ bps, $[\text{Na}^+] = 0.105$ M and $\alpha = 0.524$ °C. $T_0 = 60$ °C for both sets.

Figure 1

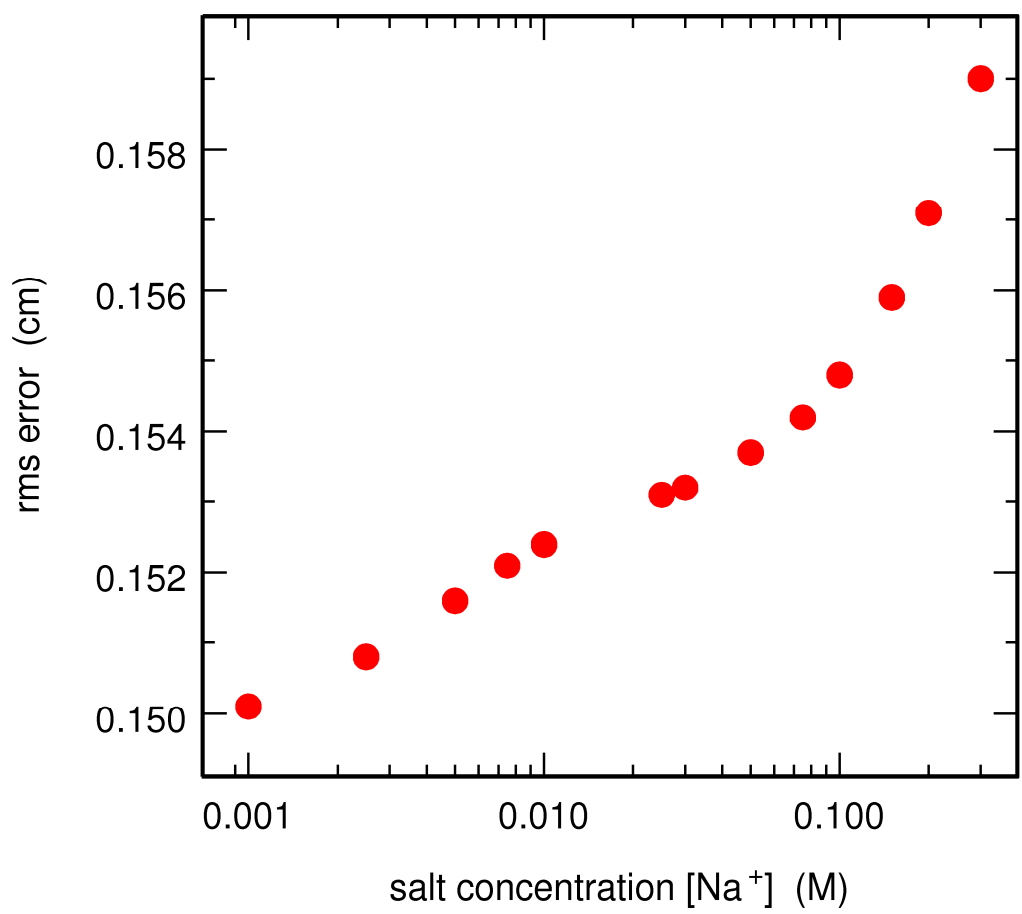

Figure 2

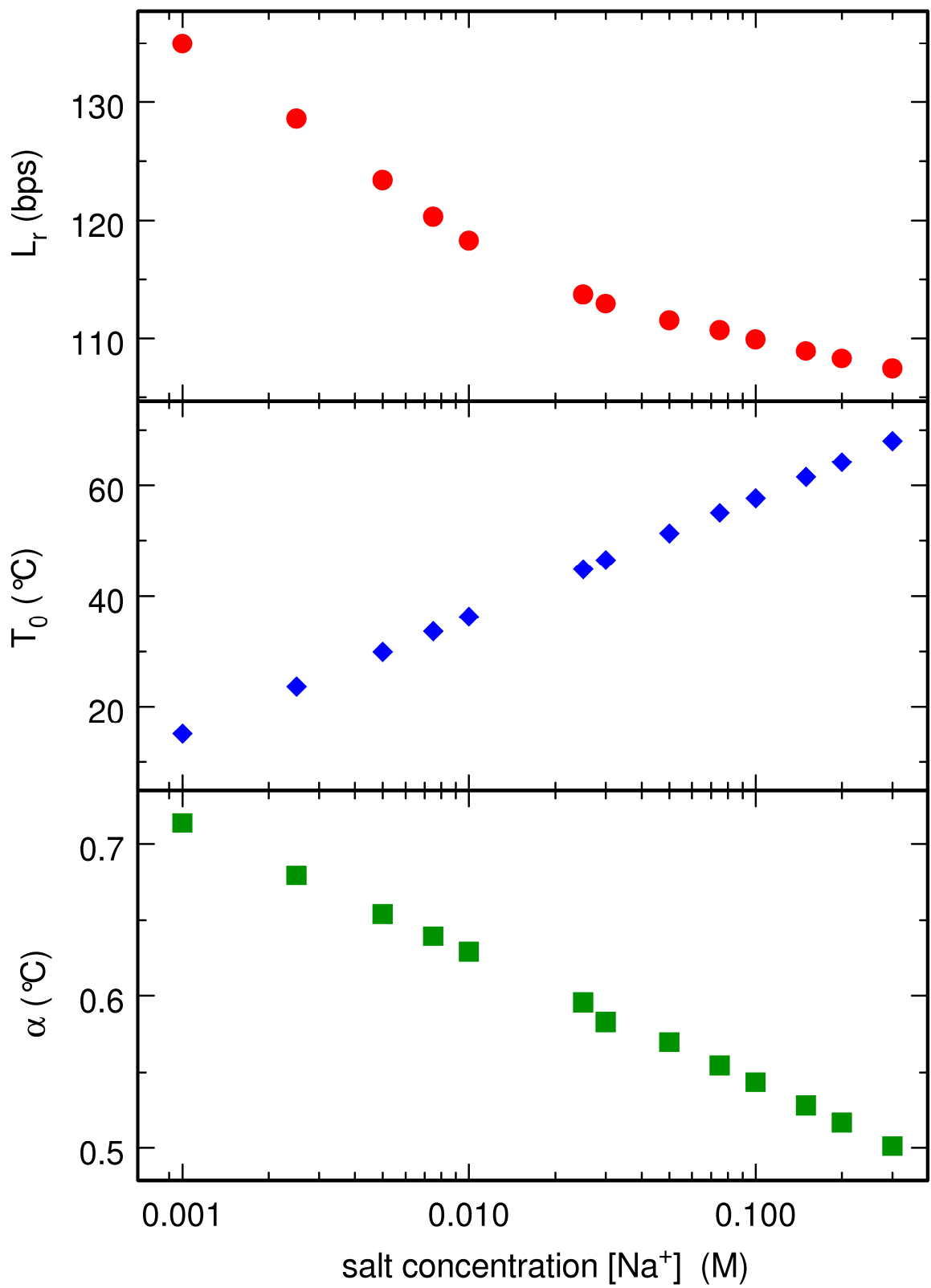

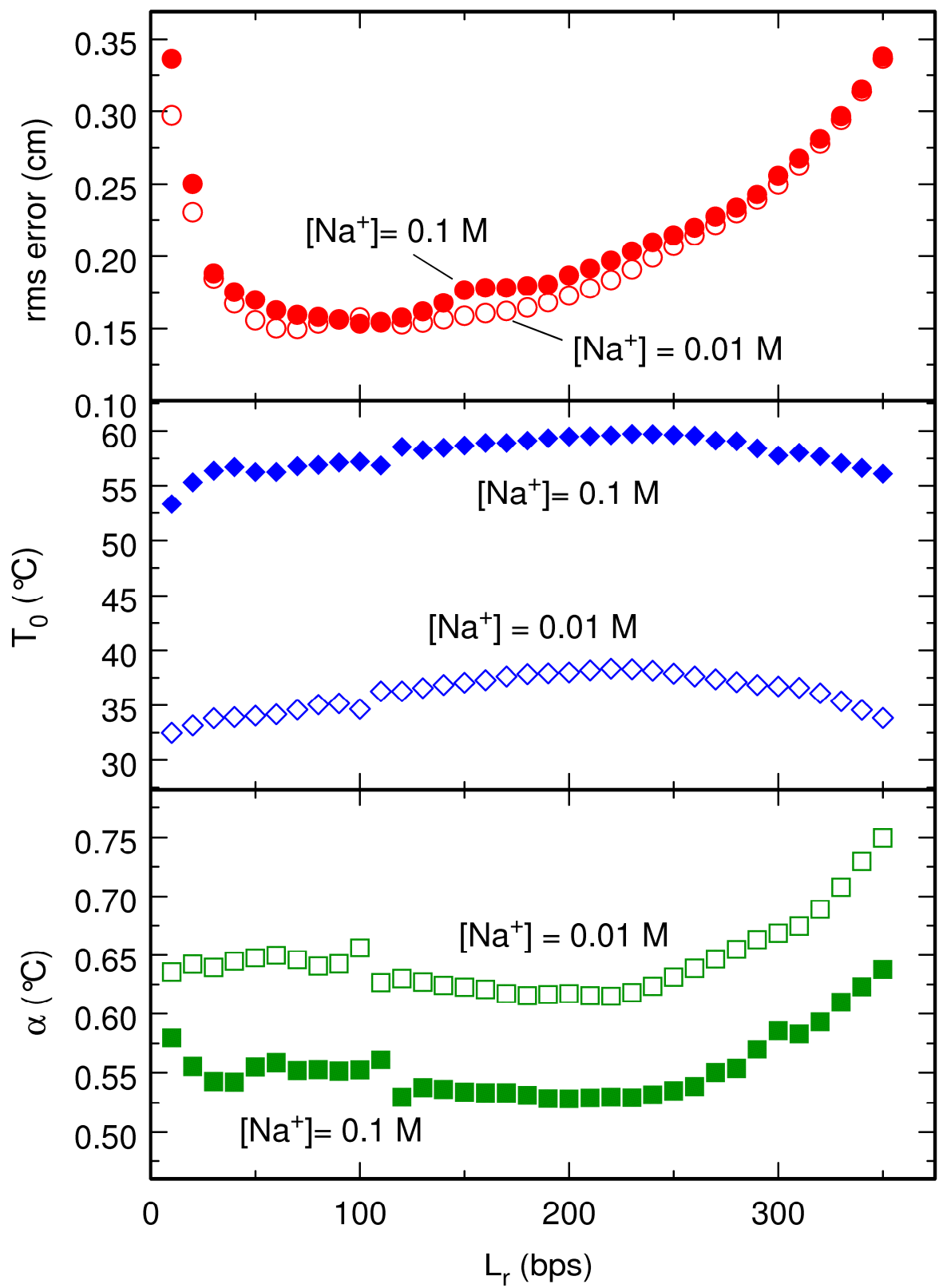

Figure 4

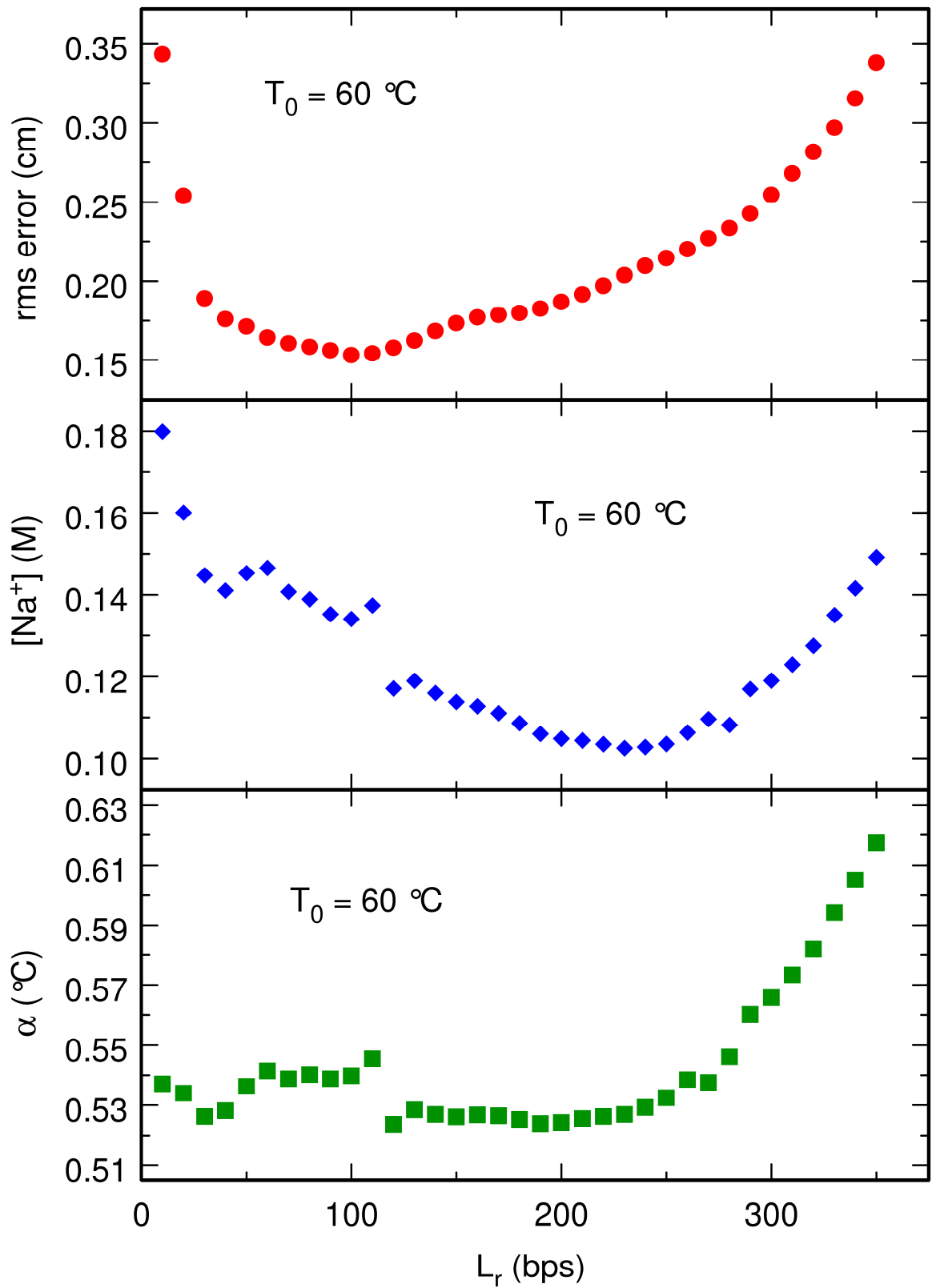
25

Figure 5

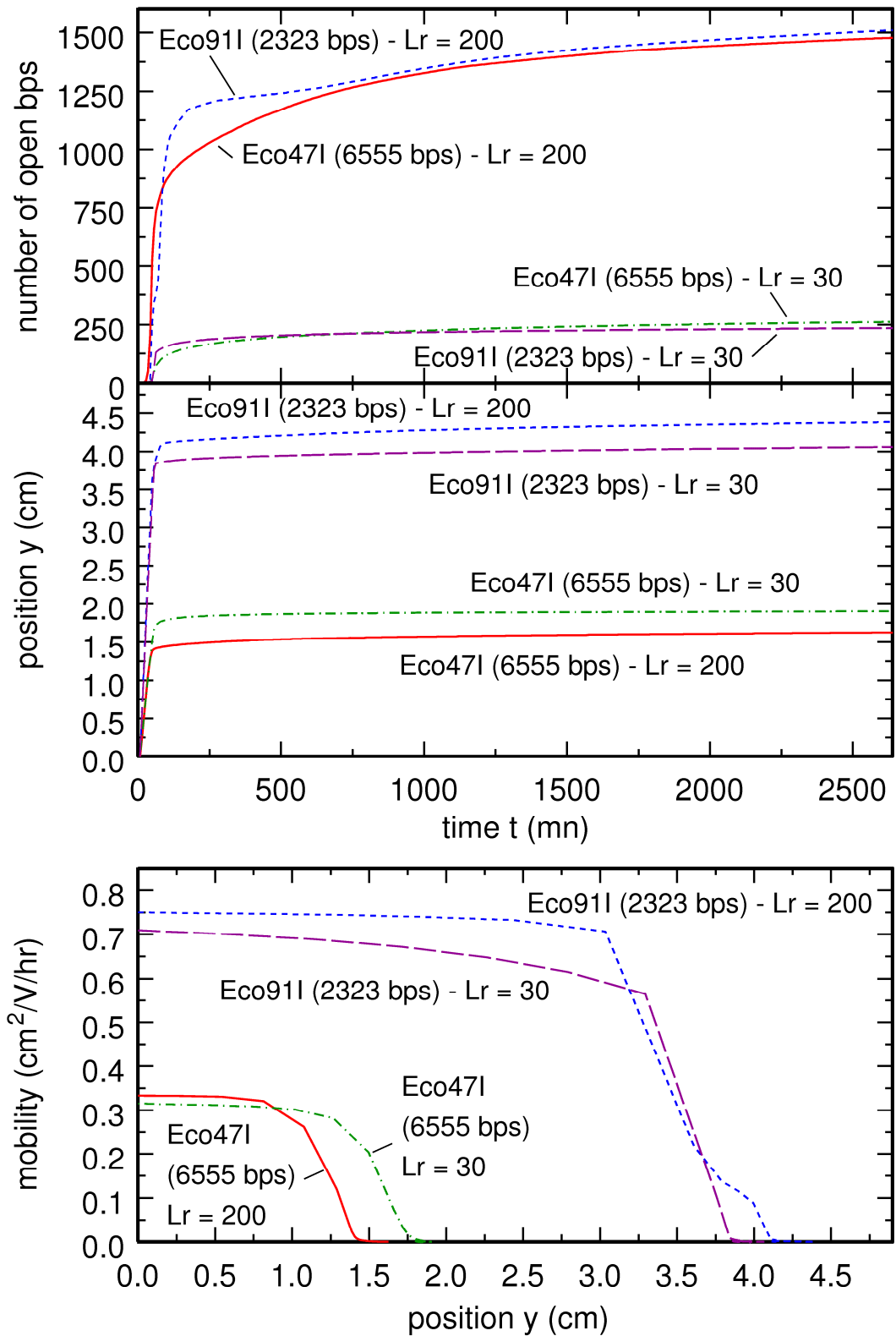